\documentstyle[aps,eqsecnum,preprint,floats,epsf,epsfig]{revtex}
\begin{document}
\def\be{\begin{eqnarray}}
\def\en{\end{eqnarray}}
\def\non{\nonumber}
\def\la{\langle}
\def\ra{\rangle}
\def\nc{N_c^{\rm eff}}
\def\vp{\varepsilon}
\def\vma{{_{V-A}}}
\def\vpa{{_{V+A}}}
\def\m{\hat{m}}
\def\ov{\overline}
\def\etapp{{\eta^{(')}}}
\def\fp{{f_{\eta'}^{(\bar cc)}}}
\def\half{{{1\over 2}}}
\def\pr{{\sl Phys. Rev.}~}
\def\prl{{\sl Phys. Rev. Lett.}~}
\def\pl{{\sl Phys. Lett.}~}
\def\np{{\sl Nucl. Phys.}~}
\def\zp{{\sl Z. Phys.}~}
\def\lsim{ {\ \lower-1.2pt\vbox{\hbox{\rlap{$<$}\lower5pt\vbox{\hbox{$\sim$}
}}}\ } }
\def\gsim{ {\ \lower-1.2pt\vbox{\hbox{\rlap{$>$}\lower5pt\vbox{\hbox{$\sim$}
}}}\ } }

\font\el=cmbx10 scaled \magstep2{\obeylines \hfill February, 2000}

\vskip 1.5 cm

\centerline{\large\bf Implications of Recent Measurements}
\centerline{\large\bf of Hadronic Charmless $B$ Decays}
\medskip
\bigskip
\centerline{\bf Hai-Yang Cheng$^1$ and Kwei-Chou Yang$^{1,2}$}
\medskip
\centerline{$^1$ Institute of Physics, Academia Sinica}
\centerline{Taipei, Taiwan 115, Republic of China}
\medskip
\centerline{$^2$ Department of Physics, Chung-Yuan Christian
University} \centerline{Chung-Li, Taiwan 320, Republic of China}
\bigskip
\bigskip
\centerline{\bf Abstract}
\bigskip
{\small Implications of recent CLEO measurements of hadronic
charmless $B$ decays are discussed. (i) Employing the
Bauer-Stech-Wirbel (BSW) model for form factors as a benchmark,
the $B\to\pi^+\pi^-$ data indicate that the form factor
$F_0^{B\pi}(0)$ is smaller than that predicted by the BSW model,
whereas the data of $B\to\omega\pi,~K^*\eta$ imply that the form
factors $A_0^{B\omega}(0),~A_0^{BK^*}(0)$ are greater than the BSW
model's values. (ii) The tree-dominated modes
$B\to\pi^+\pi^-,~\rho^0\pi^\pm,~\omega\pi^\pm$ imply that the
effective number of colors $\nc(LL)$ for $(V-A)(V-A)$ operators is
preferred to be smaller, while the current limit on $B\to\phi K$
shows that $\nc(LR)>3$. The data of $B\to K\eta'$ and $K^*\eta$
clearly indicate that $\nc(LR)\gg \nc(LL)$. (iii) In order to
understand the observed suppression of $\pi^+\pi^-$ and
non-suppression of $K\pi$ modes, both being governed by the form
factor $F_0^{B\pi}$, the unitarity angle $\gamma$ is preferred to
be greater than $90^\circ$. By contrast, the new measurement of
$B^\pm\to\rho^0\pi^\pm$ no longer strongly favors $\cos\gamma<0$.
(iv) The observed pattern $K^-\pi^+\sim \ov K^0\pi^-\sim {2\over
3}K^-\pi^0$ is consistent with the theoretical expectation: The
constructive interference between electroweak and QCD penguin
diagrams in the $K^-\pi^0$ mode explains why ${\cal B}(B^-\to
K^-\pi^0)>{1\over 2}{\cal B}(\ov B^0\to K^-\pi^+)$. (v) The
observation $\nc(LL)<3<\nc(LR)$ and our preference for
$\nc(LL)\sim 2$ and $\nc(LR)\sim 6$ are justified by a recent
perturbative QCD calculation of hadronic rare $B$ decays in the
heavy quark limit. (vi) The sizeable branching ratios of $K^*\eta$
and the enormously large decay rates of $K\eta'$ indicate that it
is the constructive interference of two comparable penguin
amplitudes rather than the mechanism specific to the $\eta'$ that
accounts for the bulk of $B\to \eta' K$ and $\eta K^*$ data. (vii)
The new upper limit set for $B^-\to\omega K^-$ no longer imposes a
serious problem to the factorization approach. It is anticipated
that ${\cal B}(B^-\to\omega K^-)\gsim 2{\cal B}(B^-\to\rho^0
K^-)\sim 2\times 10^{-6}$. (viii) An improved and refined
measurement of $B\to K^{*-}\pi^+,~\ov K^0\pi^0$ is called for in
order to resolve the discrepancy between theory and experiment.
Theoretically, it is expected that $\ov K^0\pi^0\sim{1\over
2}\,K^-\pi^0$ and $K^-\pi^+\sim 3\, K^{*-}\pi^+$.

}
\pagebreak

\section{Introduction}
A number of new hadronic charmless $B$ decay modes have been
recently reported by CLEO \cite{Gao,Kwon,Smith,Bishai,Richichi}
\be
B\to\pi^+\pi^-,~K_s^0\pi^0,~\rho^0\pi^\pm,~\omega\pi^\pm,~K^{*\pm}
\eta,~K^{*0}\eta,~\rho^\pm\pi^\mp,~K^{*\pm}\pi^\mp, \en and
several previously observed decays have received improved
measurements: $B\to K^\pm\eta',~K_s^0\eta'$,
$K^\pm\pi^\mp,~K_s^0\pi^\pm$, $K^\pm\pi^0,~\omega K^\pm$. Needless
to say, these measurements will shed light on the underlying
mechanism for charmless $B$ decays and provide important
constraints on the phenomenological models under consideration and
the parameters involved in the model, such as form factors,
unitarity angles, and nonfactorized effects.

Beyond the phenomenological level, the nonleptonic $B$ decays have
been studied within the framework of the so-called three-scale
perturbative QCD factorization theorem in which nonfactorized and
nonspectator contributions can be identified and calculated
\cite{Li}. Recently, it was shown that, in the heavy quark limit,
the hadronic matrix elements for two-body charmless $B$ decays can
be computed from first principles and expressed in terms of form
factors and meson light-cone distribution amplitudes
\cite{Beneke}. Nonfactorizable diagrams in the heavy quark limit
are dominated by hard gluon exchange and thus can be calculated as
expansion in $\alpha_s$. As we shall see below, this framework
provides a useful guidance on the nonfactorized corrections to the
hadronic matrix elements of penguin and non-penguin operators and
gives a justification on the use of generalized factorization in
which the effective Wilson coefficients $c^{\rm eff}_i$ are
renormalization-scale and -scheme independent while factorization
is applied to the tree-level hadronic matrix elements.

In the present paper we will analyze the data of hadronic
charmless $B$ decays within the framework of generalized
factorization and see what implications we can learn from the
studies of the new measured modes (1.1). This paper is organized
as follows. In Sec. II we briefly review the generalized
factorization approach relevant to rare $B$ decays. Then we
proceed to study $B\to \pi\pi$ and $\pi K$ decay modes in Sec.
III, tree-dominated modes $\rho^0\pi^\pm$ and $\omega\pi^\pm$ in
Sec. IV, $B\to K\eta',~K^*\eta$ decays in Sec. V and
$B^\pm\to\omega K^\pm$ decays in Sec. VI.  Comparison of the
present paper with the previous work \cite{CCTY} is discussed in
Sec. VII. Conclusions are presented in Sec. VIII.

\section{Framework}
In the absence of first-principles calculations for hadronic
matrix elements, it is customary to evaluate the matrix elements
under the factorization hypothesis so that $\la O(\mu)\ra$ is
factorized into the product of two matrix elements of single
currents, governed by decay constants and form factors. However,
the naive factorized amplitude is not renormalization scale- and
$\gamma_5$ scheme- independent as the scale and scheme dependence
of Wilson coefficients are not compensated by that of the
factorized hadronic matrix elements. In principle, the scale and
scheme problems with naive factorization will not occur in the
full amplitude since $\la O(\mu)\ra$ involves vertex-type and
penguin-type corrections to the hadronic matrix elements of the
4-quark operator renormalized at the scale $\mu$. Schematically,
\be
{\rm weak~decay~amplitude} &=& {\rm
naive~factorization~+~vertex\!-\!type~corrections} \\ &+&{\rm
penguin\!-\!type~corrections~+~spectator~contributions}+\cdots,
\non
\en
where the spectator contributions take into account the gluonic
interactions between the spectator quark of the $B$ meson and the
outgoing light meson. The perturbative part of vertex-type and
penguin-type corrections will render the decay amplitude scale and
scheme independent. Generally speaking, the Wilson coefficient
$c(\mu)$ takes into account the physics evolved from the scale
$M_W$ down to $\mu$, while $\la O(\mu)\ra$ involves evolution from
$\mu$ down to the infrared scale. Formally, one can write
\be
\la O(\mu)\ra=g(\mu,\mu_f)\la O(\mu_f)\ra,
\en
where $\mu_f$ is a factorization scale, and $g(\mu,\mu_f)$ is an
evolution factor running from the scale $\mu$ to $\mu_f$ which is
calculable because the infrared structure of the amplitude is
absorbed into $\la O(\mu_f)\ra$. Writing
\be
c^{\rm eff}(\mu_f)= c(\mu)g(\mu,\mu_f),
\en
the effective Wilson
coefficient will be scheme and $\mu$-scale independent. Of course,
it appears that the $\mu$-scale problem with naive factorization
is traded in by the $\mu_f$-scale problem. Nevertheless, once the
factorization scale at which we apply the factorization
approximation to matrix elements is fixed, the physical amplitude
is independent of the choice of $\mu$. More importantly, the
effective Wilson coefficients are $\gamma_5$-scheme independent.
In principle, one can work with any quark configuration, on-shell
or off-shell, to compute the full amplitude. Note that if external
quarks are off-shell and if the off-shell quark momentum is chosen
as the infrared cutoff, $g(\mu,\mu_f)$ will depend on the gauge of
the gluon field \cite{Buras98}. But this is not a problem at all
as the gauge dependence belongs to the infrared structure of the
wave function. However, if factorization is applied to $\la
O(\mu_f)\ra$, the information of the gauge dependence
characterized by the wave function will be lost. Hence, as
stressed in \cite{CLY,CCTY}, in order to apply factorization to
matrix elements and in the meantime avoid the gauge problem
connected with effective Wilson coefficients, one must work in the
on-shell scheme to obtain gauge invariant and infrared finite
$c_i^{\rm eff}$ and then applies factorization to $\la
O(\mu_f)\ra$ afterwards. Of course, physics should be $\mu_f$
independent. In the formalism of the perturbative QCD
factorization theorem, the nonperturbative meson wave functions
are specified with the dependence of the factorization scale
$\mu_f$ \cite{CLY}. These wave functions are universal for all
decay processes involving the same mesons. Hence, a consistent
evaluation of hadronic matrix elements will eventually resort to
the above-mentioned meson wave functions with $\mu_f$ dependence.

In general, the scheme- and $\mu$-scale-independent effective
Wilson coefficients have the form \cite{Ali,CT98}:
\be
\label{ceff1} c_i^{\rm eff}(\mu_f) &=& c_i(\mu)+{\alpha_s\over
4\pi}\left(\gamma_V^{T}\ln{\mu_f\over \mu}+\hat
r_V^T\right)_{ij}c_j(\mu)+~{\rm penguin\!-\!type~corrections},
\en
where $\mu_f$ is the factorization scale arising from the
dimensional regularization of infrared divergence \cite{CLY}, and
the anomalous dimension matrix $\gamma_V$ as well as the constant
matrix $\hat r_V$ arise from the vertex-type corrections to
four-quark operators. Note that in the dimensional regularization
scheme the matrix $\hat r_V$ depends on the definition of
$\gamma_5$. The infrared pole is consistently absorbed into
universal bound-state wave functions. The expressions for the {\it
gauge-invariant} constant matrix $\hat r_V$ in the naive dimension
regularization (NDR) and 't Hooft-Veltman (HV) renormalization
schemes can be found in Eqs. (2.18) and (2.19), respectively, of
\cite{CCTY}. However, the 66 and 88 entries of $\hat r_{\rm NDR}$
and $\hat r_{\rm HV}$ shown in \cite{CCTY} are erroneous: $(\hat
r_{\rm NDR})_{66}$ and $(\hat r_{\rm NDR})_{88}$ should read 17
instead of 1, while $(\hat r_{\rm HV})_{66}$ and $(\hat r_{\rm
HV})_{88}$ should read 47/3 rather than $-1/3$. This will affect
the effective Wilson coefficients $c_6^{\rm eff}$ and $c_8^{\rm
eff}$ (see Table I). For example, we have Re\,$c_6^{\rm
eff}\approx -0.060$ instead of the value $-0.048$ given in
\cite{CCTY}.

Two remarks are in order. (i) It should be stressed that the
constant matrix $\hat r_V$ arising from vertex-like corrections is
{\it not} arbitrary due to the infrared finiteness of vertex-like
diagrams: The infrared divergences in individual vertex-type
diagrams cancel in their sum. (ii) As shown in \cite{CLY}, the
evolution factor $g(\mu,\mu_f)$ can be decomposed as
$g_1(\mu)g_2(\mu_f)$, where $g_1(\mu)$ is an evolution factor from
the scale $\mu$ to $m_b$, whose anomalous dimension is the same as
that of $c(\mu)$, and $g_2(\mu_f)$ describes the evolution from
$m_b$ to $\mu_f$, whose anomalous dimension differs from that of
$c(\mu)$ because of the inclusion of the dynamics associated with
spectator quarks. Since spectator quarks do not get involved in
the factorization approximation, we will set $\mu_f=m_b$ in
practical calculations so that $g_2(\mu_f)=1$. In the generalized
factorization approach to be described below, the $\mu_f$
dependence of the effective Wilson coefficients are compensated by
that of the nonfactorized terms $\chi_i$ introduced below.

\vskip 0.4cm
\begin{table}[ht]
\caption{Numerical values of the gauge-invariant effective Wilson
coefficients $c_i^{\rm eff}$ for $b\to s$, $b\to d$ and $\bar
b\to\bar d$ transitions evaluated at $\mu_f=m_b$ and
$k^2=m_b^2/2$, where use of $|V_{ub}/V_{cb}|=0.09$ has been made.
The numerical results are insensitive to the unitarity angle
$\gamma$. \label{tab:wcs}}
\begin{center}
\begin{tabular}{ l c c c  }
 & $b\to s$,~$\bar b\to\bar s$ & $b\to d$ & $\bar b\to\bar d$  \\ \hline
$c_1^{\rm eff}$ & 1.169 & 1.169 & 1.169 \\ $c_2^{\rm eff}$ &
$-0.367$ & $-0.367$ & $-0.367$ \\ $c_3^{\rm eff}$ &
$0.0227+i0.0045$ & $0.0223+i0.0041$ & $0.0225+i0.0050$ \\
$c_4^{\rm eff}$ & $-0.0463-i0.0136$ & $-0.0450-i0.0122$ &
$-0.0458-i0.0151$
\\ $c_5^{\rm eff}$ & $0.0134+i0.0045$ & $0.0130+i0.0041$ &
$0.0132+i0.0050$ \\ $c_6^{\rm eff}$ & $-0.0600-i0.0136$ &
$-0.0588-i0.0122$ & $-0.0595-i0.0151$ \\ $c_7^{\rm eff}/\alpha$ &
$-0.0311-i0.0367$ & $-0.0286-i0.0342$ & $-0.0301-i0.0398$ \\
$c_8^{\rm eff}/\alpha$ & 0.070 & 0.070 & 0.070 \\ $c_9^{\rm
eff}/\alpha$ & $-1.429-i0.0367$ & $-1.426-i0.0342$ &
$-1.428-i0.0398$
\\ $c_{10}^{\rm eff}/\alpha$ & 0.48 & 0.48 & 0.48
\\
\end{tabular}
\end{center}
\end{table}

It is known that the effective Wilson coefficients appear in the
factorizable decay amplitudes in the combinations $a_{2i}=
{c}_{2i}^{\rm eff}+{1\over N_c}{c}_{2i-1}^{\rm eff}$ and
$a_{2i-1}= {c}_{2i-1}^{\rm eff}+{1\over N_c}{c}^{\rm eff}_{2i}$
$(i=1,\cdots,5)$. Phenomenologically, the number of colors $N_c$
is often treated as a free parameter to model the nonfactorizable
contribution to hadronic matrix elements and its value can be
extracted from the data of two-body nonleptonic decays. As shown
in \cite{Cheng94,Kamal94,Soares}, nonfactorizable effects in the
decay amplitudes of $B\to PP,~VP$ can be absorbed into the
parameters $a_i^{\rm eff}$. This amounts to replacing $N_c$ in
$a^{\rm eff}_i$ by $(N_c^{\rm eff})_i$. Explicitly,
\be
a_{2i}^{\rm eff}={c}_{2i}^{\rm eff}+{1\over (N_c^{\rm
eff})_{2i}}{c}_{2i-1}^{ \rm eff}, \qquad \quad a_{2i-1}^{\rm eff}=
{c}_{2i-1}^{\rm eff}+{1\over (N_c^{\rm eff})_{2i-1}}{c}^{\rm
eff}_{2i}, \qquad (i=1,\cdots,5),
\en
where
\be \label{nceff} (1/N_c^{\rm eff})_i\equiv
(1/N_c)+\chi_i\,,
\en
with $\chi_i$ being the nonfactorizable terms, which receive
contributions from nonfactorized vertex-type, penguin-type and
spectator corrections. In general, $\chi_i$ and $(\nc)_i$ are
complex. Recently, it has been shown in \cite{Beneke} that, in the
heavy quark limit, all nonfactorizable diagrams are dominated by
hard gluon exchange, while soft gluon effects are suppressed by
factors of $\Lambda_{\rm QCD}/m_b$. In other words, the
nonfactorized term is calculable as expansion in $\alpha_s$ in the
heavy quark limit.

To proceed, we shall assume that $\chi_i$ are universal (i.e.
process independent) in bottom decays (this amounts to assuming
generalized factorization) and that nonfactorizable effects in the
matrix elements of $(V-A)(V+A)$ operators differ from that of
$(V-A)(V-A)$ operators; that is, we shall assume that
\be
\label{chiLR} && \chi_{LL}\equiv \chi_1=\chi_2= \chi_3=
\chi_4=\chi_9=\chi_{10}, \non \\ && \chi_{LR}\equiv \chi_5=\chi_6=
\chi_7= \chi_8,
\en
and $\chi_{LR}\neq \chi_{LL}$ or equivalently
\be
&& N_c^{\rm eff}(LL)\equiv \left(N_c^{\rm
eff}\right)_1=\left(N_c^{\rm eff}\right)_2= \left(N_c^{\rm
eff}\right)_3=\left(N_c^{\rm eff}\right)_4= \left(N_c^{\rm
eff}\right)_9= \left(N_c^{\rm eff}\right)_{10},   \non\\ &&
N_c^{\rm eff}(LR)\equiv \left(N_c^{\rm
eff}\right)_5=\left(N_c^{\rm eff}\right)_6= \left(N_c^{\rm
eff}\right)_7= \left(N_c^{\rm eff}\right)_8,
\en
and $N_c^{\rm eff}(LR)\neq N_c^{\rm eff}(LL)$. As we shall see
below, the data analysis and the theoretical study of nonleptonic
rare $B$ decays all indicate that $\nc(LR)>3>\nc(LL)$. In
principle, $N_c^{\rm eff}$ can vary from channel to channel, as in
the case of charm decay. However, in the energetic two-body $B$
decays, $\nc$ is expected to be process insensitive as supported
by the data \cite{CCTY}.

Although the nonfactorized effects in hadronic charmless $B$
decays are in general small, $\chi\sim {\cal O}(0.15)$
\cite{CCTY}, they are important for the coefficients $a_2,~a_3$
and $a_5$. For example, there is a large cancellation between
$c_2^{\rm eff}$ and $c_1^{\rm eff}/N_c$, so that even a small
amount of $\chi$ will modify $a_2$ dramatically, recalling that
$a_2=c_2^{\rm eff}+c_1^{\rm eff}(1/N_c+\chi)$. Consequently, the
aforementioned coefficients are very sensitive to the change of
$\nc$, and moreover $a_2$ as well as $a_5$ have a minimum at
$\nc(LL)=\nc(LR)=3$. Therefore, nonfactorized contributions are
important to the class-II modes, e.g.
$B^0\to\pi^0\pi^0,~\rho^0\pi^0,~\omega\eta,\cdots$, and to some
decay modes which get contributions from the penguin terms
$(a_3+a_5)$, e.g. $B\to\omega K$. It is obvious that the
nonfactorized effect in these decays cannot be simply absorbed
into form factors. Another example has to do with the decays
$B\to\phi K$ and $B\to K\eta'$. In the naive factorization
approximation, the form factor $F_0^{BK}$ has to be suppressed in
order to accommodate the experimental limit on $B^-\to\phi K^-$.
However, the enormously large rate of $B\to K\eta'$ demands a
large $F_0^{BK}$. This difficulty is resolved if $\nc(LL)$ and
$\nc(LR)$ are allowed to deviate from their naive value $\nc=3$,
for example, $\nc(LL)\sim 2$ and $\nc(LR)\sim 6$ (see Sec. V).
Hence, it is inevitable to take into account nonfactorized
contributions to hadronic matrix elements in order to have a
coherent picture for rare hadronic $B$ decays.

\section{$B\to\pi\pi$ and $\pi K$ decays}
Recently CLEO has made the first observation of the decay
$B\to\pi^+\pi^-$ with the branching ratio \cite{Kwon}
\be
\label{pipi} {\cal B}(\ov B^0\to\pi^+\pi^-)
&=&(4.3^{+1.6}_{-1.4}\pm 0.5)\times 10^{-6}. \en This decay mode
puts a stringent constraint on the form factor $F_0^{B\pi}$.
Neglecting final-state interactions and employing the Wolfenstein
parameters $\rho=0.175,~\eta=0.370$, corresponding to
$\gamma\equiv{\rm Arg}(V_{ub}^*)=65^\circ$ and
$|V_{ub}/V_{cb}|=0.09$, and the effective number of colors
$\nc(LL)=2$ [see Sec. IV for a discussion of $\nc(LL)$], we find
$F_0^{B\pi}(0)=0.20\pm 0.04$.\footnote{It was argued in
\cite{Agashe} that a small value $|V_{ub}/V_{cb}|\approx 0.06$ is
preferred by the $\pi^+\pi^-$ measurement with the form factor
$F_0^{B\pi}(0)$ being fixed to be 0.33. However, this CKM matrix
element $|V_{ub}/V_{cb}|$ is smaller than the recent LEP average
$0.104^{+0.015}_{-0.018}$ \cite{LEP} and the CLEO result
 $0.083^{+0.015}_{-0.016}$ \cite{Behrens2}.} This value is
substantially smaller than the form factor $F_0^{B\pi}(0)=0.333$
obtained by Bauer, Stech and Wirbel (BSW) \cite{BSW}. This has two
important implications. First, the predicted decay rates of $B\to
K\pi$, which are mainly governed by $F_0^{B\pi}$, will in general
be smaller than the central values of experimental measurements
[see Eq. (\ref{kpi})]. Second, the form factor $F_0^{BK}(0)$ will
become smaller either. More specifically, it cannot exceed the
value, say 0.33, otherwise the SU(3)-symmetry relation
$F_0^{BK}=F_0^{B\pi}$ will be badly broken. Consequently, the
predicted $K\eta'$ rates will become too small compared to
experiment.

There are several possibilities that the $K\pi$ rates can be
enhanced : (i) The unitarity angle $\gamma$ larger than $90^\circ$
will lead to a suppresion of $B\to\pi^+\pi^-$ \cite{gamma,CCTY},
which in turn implies an enhancement of $F_0^{B\pi}$ and hence
$K\pi$ rates. (ii) A large nonzero isospin $\pi\pi$ phase shift
difference of order $70^\circ$ \cite{CCTY} can yield a substantial
suppression of the $\pi^+\pi^-$ mode. However, a large $\pi\pi$
isospin phase difference seems to be very unlikely due to the
large energy released in charmless $B$ decays. Indeed, the Regge
analysis of \cite{Gerard} indicates $\delta_{\pi\pi}=11^\circ$.
(iii) Smaller quark masses, say $m_s(m_b)=65$ MeV, will make the
$(S-P)(S+P)$ penguin terms contributing sizably to the $K\pi$
modes but less significantly to $\pi^+\pi^-$ as the penguin effect
on the latter is suppressed by the quark mixing angles.  Although
some of new quenched and unquenched lattice calculations yield
smaller $m_s$ (see e.g. \cite{Martinelli}), the value
$m_s(m_b)=65$ MeV or equivalently $m_s(1\,{\rm GeV})=100$ MeV is
barely on the verge of the lower side of lattice results
\cite{Martinelli}. Therefore, the first possibility appears to be
more plausible. Using the values $F_0^{B\pi}(0)=0.28$ and
$\gamma=105^\circ$, we find that the $\pi^+\pi^-$ decay is well
accommodated (see Table III). As a consequence, the decay rates of
$B\to K\pi$ governed by $F_0^{B\pi}$ are enhanced accordingly.

The CLEO collaboration has recently improved the measurements for
the decays $B\to K^\pm\pi^\mp,~B^\pm\to K^0\pi^\pm,~B^\pm\to
K^\pm\pi^0$ and observed for the first time the decay $\ov B^0\to
\ov K^0\pi^0$, thus completing the set of four $K\pi$ branching
ratio measurements \cite{Kwon}: \be \label{kpi}
 {\cal B}(\ov B^0\to K^-\pi^+) &=&(17.2^{+2.5}_{-2.4}\pm 1.2)\times
10^{-6}, \non
\\  {\cal B}(B^-\to \ov K^0\pi^-) &=& (18.2^{+4.6}_{-4.0}\pm 1.6)\times
10^{-6}, \non\\  {\cal B}(B^-\to K^-\pi^0) & =&
(11.6^{+3.0+1.4}_{-2.7-1.3})\times 10^{-6}, \non\\  {\cal B}(\ov
B^0\to \ov K^0\pi^0) &=& (14.6^{+5.9+2.4}_{-5.1-3.3})\times
10^{-6},
\en
which are to be compared with the 1998 results \cite{CLEOpiK}:
\be
&& {\cal B}(\ov B^0\to K^-\pi^+)=(14\pm 3\pm 2)\times 10^{-6},
\non
\\ && {\cal B}(B^-\to \ov K^0\pi^-)=(14\pm 5\pm 2)\times
10^{-6}, \non\\ && {\cal B}(B^-\to K^-\pi^0)=(15\pm 4\pm 3)\times
10^{-6}.
\en

It is known that $K\pi$ modes are penguin dominated. As far as the
QCD penguin contributions are concerned, it will be expected that
${\cal B}(\ov B^0\to K^-\pi^+)\sim {\cal B}(B^-\to\ov K^0\pi^-)$
and ${\cal B}(B^-\to K^-\pi^0)\sim {\cal B}(\ov B^0\to\ov
K^0\pi^0)\sim {1\over 2}{\cal B}(B\to K\pi^\pm)$. However, as
pointed out in \cite{CCTY,gamma}, the electroweak penguin diagram,
which can be neglected in $\ov K^0\pi^-$ and $K^-\pi^+$, does play
an essential role in the modes $K\pi^0$. With a moderate
electroweak penguin contribution, the constructive (destructive)
interference between electroweak and QCD penguins in $K^-\pi^0$
and $\ov K^0 \pi^0$ renders the former greater than the latter;
that is, ${\cal B}(B^-\to K^-\pi^0)>{1\over 2}{\cal B}(\ov
B^0\to\ov K^0\pi^-)$ and ${\cal B}(\ov B^0\to\ov K^0\pi^0)<{1\over
2}{\cal B}(\ov B^0\to K^-\pi^+)$ are anticipated. For numerical
calculations we use the parameters \be  \label{parameter} &&
m_u(m_b)=2.13\,{\rm MeV},~~~m_d(m_b)=4.27\,{\rm
MeV},~~~m_s(m_b)=85\,{\rm MeV}, \non
\\ && F_0^{B\pi}(0)=0.28,\qquad \quad
F_0^{BK}(0)=0.36,\qquad \gamma=105^\circ, \non \\
&&\nc(LL)=2,\qquad\quad \nc(LR)=6.
\en
We see from Table III that, except for the decay $\ov K^0\pi^0$,
the agreement of the calculated branching ratios for $K\pi$ modes
with experiment is excellent. By contrast, the central value of
${\cal B}(\ov B^0\to\ov K^0\pi^0)$ is much greater than the
theoretical expectation. Since its experimental error is large,
one has to await the experimental improvement to clarify the
issue. The predicted pattern
\be
K^-\pi^+\gsim \ov K^0\pi^-\sim {3\over 2}K^-\pi^0\sim 3\,\ov
K^0\pi^0
\en
is in good agreement with experiment for the first three decays.

We would like to make a remark on the trail of having
$\cos\gamma<0$. The suggestion of $\gamma>90^\circ$ or a negative
Wolfenstein parameter $\rho$ was originally motivated by the 1998
$K\pi$ data which indicated nearly equal branching ratios for the
three modes $K^-\pi^+$, $\ov K^0\pi^-$ and $K^-\pi^0$. It was
pointed out in \cite{Deshpande} that $\cos\gamma<0$ as well as a
large $m_s$, say $m_s(m_b)=200$ MeV, will allow a substantial rise
of $K^-\pi^0$ and a suppression of QCD penguin contributions so
that $K^-\pi^0\simeq K^-\pi^+$ can be accounted for. The 1999 data
\cite{Kwon} show that $K^-\pi^0\simeq {2\over 3} K^-\pi^+$, in
accordance with the theoretical anticipation. The motivation for
having a negative $\cos\gamma$ this time is somewhat different: It
provides a simply way for accommodating the suppression of
$\pi^+\pi^-$ and non-suppression of $K\pi$ data without having too
small light quark masses or too large $\pi\pi$ final-state
interactions or too small CKM matrix element $V_{ub}$.

Finally, as pointed out in \cite{CCTY}, the branching ratio of
$K^{*-}\pi^+$ predicted to be of order $0.5\times 10^{-5}$ is
smaller than that of $K^-\pi^+$ owing to the absence of the $a_6$
penguin term in the former. The observation ${\cal B}(\ov B^0\to
K^{*-}\pi^+)=(22^{+8+4}_{-6-5})\times 10^{-6}$ \cite{Gao} is thus
strongly opposite to the theoretical expectation. Clearly, it is
important to have a refined measurement of this mode.

\section{Tree-dominated charmless $B$ Decays}
CLEO has observed several tree-dominated charmless $B$ decays
which proceed at the tree level through the $b$ quark decay $b\to
u\bar ud$ and at the loop level via the $b\to d$ penguin diagrams:
$B\to \pi^+\pi^-,~\rho^0\pi^\pm,~\omega\pi^\pm,~\rho^\pm\pi^\mp$.
The first three modes have been measured recently for the first
time with the branching ratios \cite{Smith,Bishai}:
\be
\label{rhopi} {\cal B}(B^\pm\to \rho^0\pi^\pm) &=&
(10.4^{+3.3}_{-3.4}\pm2.1)\times 10^{-6}, \non\\ {\cal B}(B^\pm\to
\omega\pi^\pm) &=& (11.3^{+3.3}_{-2.9}\pm 1.5 )\times 10^{-6},
\en
and (\ref{pipi}). These decays are sensitive to the form factors
$F_0^{B\pi}$, $A_0^{B\rho}$, $A_0^{B\omega}$ and to the value of
$\nc(LL)$. To illustrate the sensitivity on form factors, we
consider two different form-factor models for heavy-to-light
transitions: the BSW model \cite{BSW} and the light-cone sum rule
(LCSR) model \cite{Ball}. The relevant form factors at zero
momentum transfer are listed in Table II. We see from Table IV
that the branching ratios of $\rho^0\pi^\pm$ and $\omega\pi^\pm$
decrease with $\nc(LL)$ and generally they become too small
compared to the data when $\nc(LL)> 3$, whereas ${\cal
B}(B\to\pi^+\pi^-)$ increases with $\nc(LL)$ and becomes too large
when $\nc(LL)>3$. More precisely, we obtain $1.1\leq \nc(LL)\leq
2.6$ from $\rho^0\pi^\pm$ and $\omega\pi^\pm$ modes. Evidently,
$\nc(LL)$ in all these tree-dominated decays are preferred to be
smaller. This is indeed what expected since the effective number
of colors, $\nc(LL)$, inferred from the Cabibbo-allowed decays
$B\to (D,D^*)(\pi,\rho)$ is in the vicinity of 2 \cite{CY} and
since the energy released in the energetic two-body charmless $B$
decays is in general slightly larger than that in $B\to D\pi$
decays, it is thus anticipated that \be \label{chi} |\chi({\rm
two-body~rare~B~decay})|\lsim |\chi(B\to D\pi)|,
\en
and hence $\nc(LL)\approx \nc(B\to D\pi)\sim 2$.

Note that the branching ratio of $\rho^0\pi^\pm$ is sensitive to
the change of the unitarity angle $\gamma$, while $\omega\pi^\pm$
is not. For example, we have ${\cal B}(B^\pm\to\rho^0\pi^\pm)\sim
{\cal B}(B^\pm\to\omega\pi^\pm)$ for $\gamma\sim 65^\circ$, and
${\cal B}(B^\pm\to\rho^0\pi^\pm)> {\cal B}(B^\pm\to\omega\pi^\pm)$
for $\gamma>90^\circ$. It appears that a unitarity angle $\gamma$
larger than $90^\circ$, which is preferred by the previous
measurement \cite{Bishai} ${\cal B}(B^\pm\to \rho^0\pi^\pm)=(15\pm
5\pm 4)\times 10^{-6}$, is no longer strongly favored by the new
data of $\rho^0\pi^\pm$.

It is worth remarking that although the decays $B\to
\rho^\pm\pi^\mp$ are sensitive to $\nc(LL)$, no useful constraint
can be extracted at this moment from the present measurement
\cite{Smith}: ${\cal B}(\ov
B^0\to\rho^+\pi^-+\rho^-\pi^+)=(27.6^{+8.4}_{-7.4}\pm 4.2)\times
10^{-6}$ due to its large error.

From Tables II and IV it is also clear that the form factor
$A_0^{B\omega}$ predicted by the LCSR, which are substantially
larger than that of the BSW model, is more favored. Since the form
factor $F_0^{B\pi}$ in the LCSR is slightly big, we will employ
the improved LCSR model (called as LCSR$'$), which is the same as
the LCSR of \cite{Ball} except that the values of $F_0^{B\pi}(0)$
and $F_0^{BK}(0)$ are replaced by those given in
(\ref{parameter}), in ensuing calculations.

\begin{table}[ht]
{\small Table II.~Form factors at zero momentum transfer for $B\to
P$ and $B\to V$ transitions evaluated in the BSW model \cite{BSW}.
The values given in the square brackets are obtained in the
light-cone sum rule (LCSR) analysis \cite{Ball}. We have assumed
SU(3) symmetry for the $B\to\omega$ form factors in the LCSR
approach. In realistic calculations we use Eq. (3.13) of
\cite{CCTY} for $B\to\etapp$ form factors. For later purposes, we
will use the improved LCSR model (LCSR$'$) for form factors, which
is the same as the LCSR of \cite{Ball} except for the values of
$F_0^{B\pi}(0)$ and $F_0^{BK}(0)$ being replaced by those given in
Eq. (\ref{parameter}).}
\begin{center}
\begin{tabular}{ l l c c c c }
Decay  & $F_1=F_0$ & $V$ & $A_1$ & $A_2$ & $A_3=A_0$ \\ \hline
$B\to\pi^\pm$ & 0.333~[0.305] & & & &  \\ $B\to K$ & 0.379~[0.341]
& & & & \\ $B\to\eta$ & 0.168~[---] & & & & \\ $B\to\eta'$ &
0.114~[---] & & & & \\ $B\to\rho^\pm$ & & 0.329~[0.338] &
0.283~[0.261] & 0.283~[0.223] & 0.281~[0.372] \\ $B\to\omega$ & &
0.232~[0.239] & 0.199~[0.185] & 0.199~[0.158] & 0.198~[0.263] \\
$B\to K^*$ & & 0.369~[0.458] & 0.328~[0.337] & 0.331~[0.283] &
0.321~[0.470] \\
\end{tabular}
\end{center}
\end{table}

\section{$B\to K\eta'$ and $K^*\eta$ decays}
The improved measurements of the decays $B\to\eta' K$ by CLEO
\cite{Richichi} \be  {\cal B}(B^\pm\to\eta' K^\pm) &=&
\left(80^{+10}_{-~9}\pm 7\right)\times 10^{-6}, \non \\
 {\cal B}(B^0\to\eta' K^0) &=& \left(89^{+18}_{-16}\pm 9
\right)\times 10^{-6},
\en
are larger than the previous published results \cite{Behrens}:
\be
\label{Ketap} {\cal B}(B^\pm\to\eta' K^\pm) &=&
\left(65^{+15}_{-14}\pm 9\right)\times 10^{-6}, \non \\
 {\cal B}(B^0\to\eta' K^0) &=& \left(47^{+27}_{-20}\pm 9
\right)\times 10^{-6}.
\en
This year CLEO has also reported the new measurement of $B\to
K^*\eta$ with the branching ratios \cite{Richichi} \be  {\cal
B}(B^\pm\to\eta K^{*\pm}) &=& \left(26.4^{+9.6}_{-8.2}\pm
3.3\right)\times 10^{-6}, \non \\
 {\cal B}(B^0\to\eta K^{*0}) &=& \left(13.8^{+5.5}_{-4.4}\pm 1.6
\right)\times 10^{-6}.
\en

Theoretically, the branching ratios of $K\eta'$ ($K^*\eta$) are
anticipated to be much greater than $K\pi$ ($K^*\eta'$) modes
owing to the presence of constructive interference between two
penguin amplitudes arising from non-strange and strange quarks of
the $\eta'$ or $\eta$.\footnote{In a recent analysis \cite{Hou},
the branching ratio of $K^*\eta'$ is predicted to be similar to
that of $K^*\eta$, whereas it is found not exceed $1\times
10^{-6}$ according to \cite{Ali} and the present paper.} In
general, the decay rates of $K\eta'$ increase slowly with
$\nc(LR)$ if $\nc(LL)$ is treated to be the same as $\nc(LR)$, but
fast enough with $\nc(LR)$ if $\nc(LL)$ is fixed at the value of
2. Evidently, the data much favor the latter case (see Table
III).\footnote{As stressed in \cite{CCTY}, the contribution from
the $\eta'$ charm content will make the theoretical prediction
even worse at the small values of $1/\nc$ if $\nc(LL)=\nc(LR)$\,!
On the contrary, if $\nc(LL)\approx 2$, the $c\bar c$ admixture in
the $\eta'$ will always lead to a constructive interference
irrespective of the value of $\nc(LR)$.} On the contrary, the
branching ratios of $K^*\eta$ in general decrease with $\nc(LR)$
when $\nc(LL)=\nc(LR)$ but increase with $\nc(LR)$ when
$\nc(LL)=2$. Again, the latter is preferred by experiment. Hence,
the data of both $K\eta'$ and $K^*\eta$ provide another strong
support for a small $\nc(LL)$ and for the relation
$\nc(LR)>\nc(LL)$. In other words, the nonfactorized effects due
to $(V-A)(V-A)$ and $(V-A)(V+A)$ operators should be treated
differently.

It appears from Tables III and IV that the data of $K^*\eta$ and
in particular $K\eta'$ are well accommodated by $\nc(LR)=\infty$.
However, we have argued in \cite{CCTY} that $\nc(LR)\lsim 6$. In
principle, the value of $\nc(LR)$ can be extracted from the decays
$B\to\phi K$ and $\phi K^*$. The present limit \cite{Gao,Bishai}
\be \label{phiK} {\cal B}(B^\pm\to\phi K^\pm)< 0.59\times 10^{-5}
\en
at 90\% C.L. implies that \be \label{ncLR} \nc(LR)\geq \cases{5.0
& BSW, \cr 4.2 & LCSR$'$,  \cr}
\en
with $\nc(LL)$ being fixed at the value of 2. Note that this
constraint is subject to the corrections from spacelike penguin
and $W$-annihilation contributions. At any rate, it is safe to
conclude that $\nc(LR)>3>\nc(LL)$.

Since the penguin matrix elements of scalar and pseudoscalar
densities are sensitive to the strange quark mass, the discrepancy
between theory and experiment, especially for $K\eta'$, can be
further improved by using an even smaller $m_s$, say $m_s(m_b)=65$
MeV. However, as remarked in Sec. III, this small strange quark
mass is not favored by lattice calculations. Moreover, it will
lead to too large $B\to K\pi$ rates. For example, the predicted
${\cal B}(\ov B^0\to K^-\pi^+)=28\times 10^{-6}$ using
$m_s(m_b)=65$ MeV is too large compared to the observed branching
ratio $(17.2^{+2.5}_{-2.4}\pm 1.2)\times 10^{-6}$.

Several new mechanisms have been proposed in the past few years to
explain the observed enormously large rate of $K\eta'$, for
example, the large charm content of the $\eta'$ \cite{Halperin} or
the two-gluon fusion mechanism via the anomaly coupling of the
$\eta'$ with two gluons \cite{Ahmady,Du}. These mechanisms will in
general predict a large rate for $K^*\eta'$ comparable to or even
greater than $K\eta'$ and a very small rate for $K^*\eta$ and
$K\eta$. The fact that the $K^*\eta$ modes are observed with
sizeable branching ratios indicates that it is the constructive
interference of two comparable penguin amplitudes rather than the
mechanism specific to the $\eta'$ that accounts for the bulk of
$B\to \eta' K$ and $\eta K^*$ branching ratios.

Two remarks are in order. First, as shown in \cite{CCTY}, the
charged $\eta' K^-$ mode gets enhanced when $\cos\gamma$ becomes
negative while the neutral $\eta' K^0$ mode remains steady.
Therefore, it is important to see if the disparity between $\eta'
K^\pm$ and $\eta' K^0$ is confirmed when experimental errors are
improved and refined in the future. Second, we see from Table IV
that the form factor $A_0^{BK^*}(0)$ entering the decay amplitude
of $B\to K^*\eta$ is preferred to be larger than the value
predicted by the BSW model.

The observation $\nc(LL)<3<\nc(LR)$ is theoretically justified by
a recent perturbative QCD calculation of $B\to\pi\pi$ decays in
the heavy quark limit. Following the notation of \cite{Beneke}, we
find the nonfactorized terms:\footnote{Note that Eqs. (4-8) in
\cite{Beneke} can be reproduced from Eqs. (2.12-2.19) and (4.1) in
\cite{CCTY} with the nonfactorized terms given by Eq.
(\ref{nonf}). For example, from \cite{CCTY} we obtain in the NDR
scheme (the superscript ``eff" of $a_i$ is dropped for
convenience) that \be \label{a24}a_2 &=& c_2+{c_1\over
N_c}+{\alpha_s\over 4\pi}\,{C_F\over N_c}c_1\,(12\ln{m_b\over
\mu}-18)+c_1^{\rm eff}\chi_2, \non \\ a_4 &=& c_4+{c_3\over
N_c}+{\alpha_s\over 4\pi}\,{C_F\over N_c}c_3\,(12\ln{m_b\over
\mu}-18)    +c_3^{\rm eff}\chi_4 +{\alpha_s\over
9\pi}(C_t+C_p+C_g), \non \\ a_5 &=& c_5+{c_6\over
N_c}+{\alpha_s\over 4\pi}\,{C_F\over N_c}c_6(-12\ln{m_b\over
\mu}+6)+c_6^{\rm eff}\chi_5,
\en
with $C_t,~C_p,~C_g$ being defined in \cite{Ali} and
$C_F=(N_c^2-1)/(2N_c)$, while Eqs. (6) and (8) of \cite{Beneke}
lead to
\be \label{a24B} a_2 &=& c_2+{c_1\over N_c}+{\alpha_s\over
4\pi}\,{C_F\over N_c}c_1\,(12\ln{m_b\over \mu}-18+f^{\rm I}+f^{\rm
II}), \non
\\ a_4 &=& c_4+{c_3\over N_c}+{\alpha_s\over 4\pi}\,{C_F\over
N_c}c_3\,(12\ln{m_b\over \mu}-18+f^{\rm I}+f^{\rm II})
+{\alpha_s\over 9\pi}(C_t+C_p+C_g), \non\\ a_5 &=& c_5+{c_6\over
N_c}+{\alpha_s\over 4\pi}\,{C_F\over N_c}c_6(-12\ln{m_b\over
\mu}+6-f^{\rm I}-f^{\rm II}),
\en
where the hard scattering function $f^{\rm I}$ corresponds to hard
gluon exchange between the two outgoing light mesons and $f^{\rm
II}$ describes the hard nonfactorized effect involving the
spectator quark of the $B$ meson. The expressions for the hard
scattering functions $f^{\rm I}$ and $f^{\rm II}$ can be found in
\cite{Beneke}. Comparing (\ref{a24}) with (\ref{a24B}) yields
\be
\chi_2=\chi_4=-\chi_5={\alpha_s\over 4\pi}\,{C_F\over N_c}(f^{\rm
I}+f^{\rm II}). \non
\en
Note that the quark mass entering into the penguin matrix elements
of scalar and pseudoscalar densities via equations of motion is
fixed at the scale $\mu_f$. } \be \label{nonf}
\chi_1=\chi_2=\chi_3=\chi_4=-\chi_5=-\chi_6={\alpha_s\over
4\pi}\,{C_F\over N_c}(f^{\rm I}+f^{\rm II}).  \en
It follows from
(\ref{chiLR}) that
\be
\chi_{LR}=-\chi_{LL}=-{\alpha_s\over 4\pi}\,{C_F\over N_c}(f^{\rm
I}+f^{\rm II}).
\en
Several remarks are in order. (i) Since $f^{\rm I}$ is complex due
to final-state interactions via hard gluon exchnge \cite{Beneke},
so are $\chi_i$ and $\nc(LL)$ and $\nc(LR)$. Nevertheless, the
complex phases of $\chi_i$ are in general small. (ii) Contrary to
the common assertion, the nonfactorized term is dominated by hard
gluon exchange in the heavy quark limit as soft gluon
contributions to $\chi_i$ are suppressed by orders of
$\Lambda_{\rm QCD}/m_b$ \cite{Beneke}. (iii) Because
Re\,$\chi_{LL}>0$, it is obvious that $\nc(LL)<3$ and $\nc(LR)>3$
[see Eq. (\ref{nceff})]. Furthermore, $\nc(LL)\sim 2$ implies
$\nc(LR)\sim 6$. Therefore, the assumption (\ref{chiLR}) and the
empirical observation $\nc(LR)>3>\nc(LL)$ get a  justification
from the perturbative QCD calculation performed in the heavy quark
limit.

\section{ $B\to\omega K$ and $\rho K$ decays}
The previous CLEO observation \cite{CLEOomega} of a large
branching ratio for $B^\pm\to \omega K^\pm$
\be \label{omegaK}
{\cal B}(B^\pm\to\omega K^\pm)=\left(15^{+7}_{-6}\pm 2\right)
\times 10^{-6},
\en
imposes a serious problem to the generalized factorization
approach: The observed rate is enormously large compared to naive
expectation \cite{CCTY}. Since the $\omega K^-$ amplitude differs
from that of $\rho^0 K^-$ only in the QCD penguin term
proportional to $(a_3+a_5)$ and in the electroweak penguin term
governed by $a_9$, it is naively anticipated that their branching
ratios are similar as the contributions from $a_3,a_5,a_9$ are not
expected to be large. While the branching ratio of $B^\pm\to\rho^0
K^\pm$ is estimated to be of order $1\times 10^{-6}$ (see Table
IV), the prediction of ${\cal B}(B^\pm\to\omega K^\pm)$ is less
certain because the penguin contribution proportional to
$(a_3+a_5)$ depends sensitively on $\nc(LR)$. At any rate, it is
reasonable to assert that ${\cal B}(B^-\to\omega K^-)\gsim 2{\cal
B}(B^-\to\rho^0 K^-)\sim 2\times 10^{-6}$.

As pointed out recently in \cite{Bishai}, the additional data and
re-analysis of old CLEO data did not support the previously
reported observation (\ref{omegaK}). Therefore, the new
measurement of $B^-\to\omega K^-$ no longer imposes a serious
difficulty to the factorization approach. The theoretical
prediction ${\cal B}(B^-\to\omega K^-)\sim (2.1-5.5)\times
10^{-6}$ for $\nc(LL)=2$ and $\nc(LR)$ ranging from 6 to $\infty$
is consistent with the current limit $8.0\times 10^{-6}$
\cite{Bishai}.  It is important to measure the branching ratios of
$\omega K$ and $\rho K$ modes in order to understand their
underlying mechanism. From Table IV we see that $\rho^0 K^0\sim
\rho^+ K^0>\rho^0 K^+$ is expected in the factorization approach.

\section{Comparison with Ref. [7]}
Although we have followed the framework of \cite{CCTY} to study
nonleptonic charmless $B$ decays, it is useful at this point to
summarize the differences between the present work and
\cite{CCTY}:
\begin{itemize}
\item The 66 and 88 entries of the constant matrix $\hat r_V$ in
NDR and HV $\gamma_5$ schemes given in \cite{CCTY} are erroneous
and have been corrected here. As a result, the magnitude of the
effective penguin Wilson coefficient $c_6^{\rm eff}$ is enhanced.
The decay rates of the penguin-dominated modes governed by the
$a_6$ penguin term are thus enhanced. For example, the branching
ratios of $\ov K^{*0}\eta$ and $K^{*-}\eta$ are enhanced by almost
a factor of 2.
\item In order to accommodate the new data of $\pi^+\pi^-$ and
$K\pi$ decays, we have fixed the relevant form factors and the
unitarity angle to be $F_0^{B\pi}(0)=0.28,~F_0^{BK}(0)=0.36$ and
$\gamma=105^\circ$.
\item While the strange quark mass is slightly changed to
$m_s(m_b)=85\,{\rm MeV}$, the $u$ and $d$ quark masses are
modified to $m_u(m_b)=2.13\,{\rm MeV},~m_d(m_b)=4.27\,{\rm MeV}$
in order to respect the chiral-symmetry relation
$m^2_{\pi^\pm}/(m_u+m_d)=m^2_{K^+}/(m_u+m_s)$.
\item Branching ratios of all $B_{u,d}\to PP,VP,VV$ modes are
tabulated in Tables III-V in BSW and LCSR$'$ models for form
factors. Our preference for heavy-to-light form factors is that
given by the LCSR$'$ model.
\end{itemize}

\section{Conclusions}
Implications inferred from recent CLEO measurements of hadronic
charmless two-body decays of $B$ mesons are discussed in the
present paper. Our main conclusions are as follows.
\begin{enumerate}
\item Employing the Bauer-Stech-Wirbel (BSW) model for form
factors as a benchmark, the $B\to\pi^+\pi^-$ data indicate that
the form factor $F_0^{B\pi}(0)=0.20\pm0.04$ is much smaller than
that predicted by the BSW model, whereas the data of
$B\to\omega\pi,~K^*\eta$ imply that the form factors
$A_0^{B\omega}(0),~A_0^{BK^*}(0)$ are greater than the values
obtained in the BSW model.

\item The tree-dominated modes
$B\to\pi^+\pi^-,~\rho^0\pi^\pm,~\omega\pi^\pm$ imply that the
effective number of colors $\nc(LL)$ for $(V-A)(V-A)$ operators is
preferred to be smaller $1.1\leq \nc(LL)\leq 2.6$, while the
current limit on $B\to\phi K$ shows that $\nc(LR)>3$. The data of
$B\to K\eta'$ and $K^*\eta$ clearly support the observation
$\nc(LR)\gg \nc(LL)$.

\item The decay rates of $\pi^+\pi^-$ and
$K\pi$ are governed by the form factor $F_0^{B\pi}$. In order to
explain the observed suppression of $\pi^+\pi^-$ and
non-suppression of $K\pi$ modes, the unitarity angle $\gamma$ is
favored to be greater than $90^\circ$ (see also
\cite{Hou,Gronau}). By contrast, the new measurement of
$B^\pm\to\rho^0\pi^\pm$ no longer strongly favors $\cos\gamma<0$.

\item The observed pattern $K^-\pi^+\sim \ov K^0\pi^-\sim {3\over
2}K^-\pi^0$ is consistent with the theoretical expectation: The
constructive interference between electroweak and QCD penguin
diagrams in the $K^-\pi^0$ mode explains why ${\cal B}(B^-\to
K^-\pi^0)>{1\over 2}{\cal B}(\ov B^0\to K^-\pi^+)$.

\item We found that, except for the decays $K^{*-}\pi^+$ and $\bar
K^0\pi^0$, all the measured charmless $B$ decays can be well
accommodated by the LCSR$'$ form factors and the parameters
$m_s(m_b)=85\,{\rm MeV},~F_0^{B\pi}(0)=0.28,~F_0^{BK}(0)=0.36,\
\gamma=105^\circ,~\nc(LL)=2,~\nc(LR)=6$.

\item The observation $\nc(LL)<3<\nc(LR)$ and our preference for
$\nc(LL)\sim 2$ and $\nc(LR)\sim 6$ are theoretically justified by
a recent perturbative QCD calculation of charmless $B$ decays
performed in the heavy quark limit.

\item The new upper limit set for $B^-\to\omega K^-$ no
longer imposes a serious problem to the factorization approach. It
is anticipated that ${\cal B}(B^-\to\omega K^-)\gsim 2{\cal
B}(B^-\to\rho^0 K^-)\sim 2\times 10^{-6}$.

\item An improved and refined measurement of $B\to K^{*-}\pi^+,~\ov
K^0\pi^0$ is called for in order to resolve the discrepancy
between theory and experiment. Theoretically, it is expected that
$\ov K^0\pi^0\sim{1\over 2}\,K^-\pi^0$ and $K^-\pi^+\sim 3\,
K^{*-}\pi^+$.

\item Theoretical calculations suggest that the following 18 decay
modes of $B^-_u$ and $\ov B^0_d$ have branching ratios are of
order $10^{-5}$ or above (in sequence of their decay rate
strength): $\eta'K^-,~\eta'K^0$, $\rho^+\rho^-,~\rho^-\rho^0$,
$\rho^- \omega,~K^-\pi^+$, $\bar K^0\pi^-,~\rho^-\pi^+$,
$K^{*-}\eta$, $\rho^+\pi^-,~\rho^-\pi^0$,
$K^-\pi^0,~\rho^0\pi^-,~\ov K^{*0}\eta$,
$\omega\pi^-,~K^{*-}\rho^0$, $K^{*-}\rho^+,~\rho^-\eta$. Many of
them have been observed and the rest will have a good chance to be
seen soon.

\end{enumerate}

\acknowledgements  This work is supported in part by the National
Science Council of the Republic of China under Grant No.
NSC89-2112-M001-016.

%%%%%%%%%%%%%%%%%%%%%%%%%%%%%%%%%%%%%%%%%%%%%%%%%%%%%%%%

{\squeezetable
\begin{table}[ht]
{\footnotesize Table III. Branching ratios (in units of $10^{-6}$)
averaged over CP-conjugate modes for charmless $B_{u,d}\to PP$
decays. Predictions are made for $k^2=m_b^2/2$,
$\sqrt{\rho^2+\eta^2}=0.41,~\gamma=105^\circ$, and
$\nc(LR)=2,3,6,\infty$ with $\nc(LL)$ being fixed to be 2 in the
first case and treated to be the same as $\nc(LR)$ in the second
case. Classification of decay amplitudes is described in details
in \cite{CCTY}. Results using the improved light-cone sum rule
(LCSR$'$) and the BSW model for heavy-to-light form factors are
shown in the upper and lower entries, respectively. Experimental
values (in units of $10^{-6}$) are taken from
\cite{Gao,Bishai,Smith,Richichi,Kwon,Behrens}. Our preferred
predictions for branching ratios are those using LCSR$'$ form
factors, $\nc(LL)=2$ and $\nc(LR)=6$.}
\begin{center}
\begin{tabular}{l l c c c c c c c c l}
 &  & \multicolumn{4}{c}{$\nc(LL)=2$}
 &   \multicolumn{4}{c}{$\nc(LL)=\nc(LR)$}  \\ \cline{3-6} \cline{7-10}
\raisebox{2.0ex}[0cm][0cm]{Decay} &
\raisebox{2.0ex}[0cm][0cm]{Class} & 2 & 3 & 6 & $\infty$ & 2 & 3 &
6 & $\infty$  & \raisebox{2.0ex}[0cm][0cm]{Expt.}\\ \hline
 $\ov B^0_d \to \pi^+ \pi^-$&        I
  &5.98&5.95&5.92&5.90&5.98&6.79&7.69&8.58& $ 4.3^{+1.6}_{-1.4}
  \pm 0.5$ \\
 &&8.31&8.27&8.23&8.20&8.31&9.44&10.6&11.9& \\
 $\ov B^0_d \to \pi^0\pi^0$ &       II,VI
  &0.54&0.56&0.58&0.60&0.54&0.22&0.18&0.43&$<9.3$\\
 &&0.75&0.78&0.80&0.83&0.75&0.31&0.26&0.60& \\
 $\ov B^0_d \to \eta \eta$  &       II,VI
  &0.13&0.14&0.16&0.18&0.13&0.18&0.37&0.69&$<18$\\
 &&0.18&0.20&0.23&0.26&0.18&0.25&0.52&0.98& \\
 $\ov B^0_d \to \eta \eta'$ &       II,VI
  &0.11&0.16&0.24&0.34&0.11&0.12&0.32&0.69& $<27$\\
 &&0.16&0.23&0.34&0.48&0.16&0.18&0.45&0.98& \\
 $\ov B^0_d \to \eta' \eta'$&       II,VI
  &0.03&0.05&0.09&0.14&0.03&0.02&0.06&0.17&$<47$\\
 &&0.04&0.07&0.12&0.21&0.04&0.02&0.09&0.24& \\
  $ B^- \to \pi^- \pi^0$ &          III
  &5.92&5.92&5.92&5.92&5.92&4.68&3.58&2.64&$< 12.7$\\
 &&8.23&8.23&8.23&8.24&8.23&6.50&4.98&3.67& \\
 $ B^- \to \pi^- \eta$  &           III
  &2.64&2.65&2.66&2.68&2.64&2.05&1.60&1.30& $<5.7$\\
 &&3.68&3.69&3.71&3.73&3.68&2.85&2.22&1.80& \\
 $ B^- \to \pi^-\eta'$  &           III
  &1.79&1.75&1.77&1.86&1.79&1.30&0.91&0.62& $<12$\\
 &&2.51&2.45&2.48&2.62&2.51&1.81&1.26&0.85& \\
 $\ov B^0_d \to   K^- \pi^+$&       IV
  &16.9&17.7&18.6&19.5&16.9&18.5&20.2&22.0&$
  17.2^{+2.5}_{-2.4}\pm1.2$\\
 &&23.7&24.9&26.1&27.4&23.7&26.0&28.4&30.9& \\
 $ B^-\to \ov K^0\pi^-$       &            IV
  &15.2&16.1&17.0&17.9&15.2&17.5&20.0&22.6&
  $18.2^{+4.6}_{-4.0}\pm 1.6 $\\
 &&21.4&22.6&23.8&25.1&21.4&24.6&28.0&31.7& \\
 $ B^- \to K^-   K^0$   &                   IV
  &1.76&1.86&1.97&2.07&1.76&2.02&2.31&2.61&$ <5.1$\\
 &&1.97&2.08&2.19&2.31&1.97&2.26&2.57&2.91& \\
 $\ov B^0_d\to\ov K^0\pi^0$&        VI
  &5.27&5.63&6.02&6.41&5.27&6.20&7.25&8.39&
  $14.6^{+5.9+2.4}_{-5.1-3.3}$\\
 &&7.66&8.18&8.73&9.28&7.66&9.00&10.5&12.1& \\
 $ \ov B^0_d\to K^0\ov K^0$ &             VI
  &1.65&1.75&1.85&1.95&1.65&1.90&2.17&2.45&$ <17$\\
 &&1.85&1.95&2.06&2.17&1.85&2.12&2.42&2.73& \\
 $\ov B^0_d \to \pi^0 \eta$ &       VI
  &0.36&0.40&0.45&0.50&0.36&0.41&0.47&0.54& $<2.9$\\
 &&0.50&0.56&0.63&0.69&0.50&0.58&0.66&0.75& \\
 $\ov B^0_d \to \pi^0 \eta'$&       VI
  &0.11&0.20&0.33&0.48&0.11&0.12&0.14&0.17& $<5.7$\\
 &&0.15&0.29&0.46&0.69&0.15&0.17&0.20&0.24& \\
 $\ov B^0_d \to\ov K^0\eta$&       VI
  &1.74&1.57&1.40&1.24&1.74&2.38&3.13&4.01&$<9.3 $\\
 &&1.30&1.12&0.96&0.81&1.30&1.83&2.45&3.26& \\
 $\ov B^0_d\to\ov K^0\eta'$&       VI
  &25.8&36.0&47.9&61.6&25.8&27.9&30.1&32.4&
 $ 89^{+18}_{-16}\pm9 $\\
 &&31.4&43.4&57.5&73.5&31.4&34.0&36.8&39.6& \\
 $ B^-\to K^- \pi^0$    &           VI
  &11.6&12.1&12.6&13.2&11.6&12.4&13.3&14.2&
  $11.6^{+3.0+1.4}_{-2.7-1.3} $\\
 &&15.9&16.6&17.3&18.0&15.9&17.0&18.2&19.5& \\
 $ B^- \to K^- \eta$    &           VI
  &1.75&1.59&1.43&1.28&1.75&2.44&3.29&4.32&$<6.9$\\
 &&1.37&1.21&1.07&0.94&1.37&1.93&2.67&3.58& \\
 $ B^- \to K^- \eta'$   &           VI
  &28.8&39.9&52.8&67.5&28.8&30.6&32.5&34.5&
 $80^{+10}_{-~9}\pm7$\\
 &&35.2&48.3&63.5&80.8&35.2&37.5&39.8&42.3& \\
\end{tabular}
\end{center}
\end{table}
\vskip 0.4cm
 }

\begin{table}[ht]
{\footnotesize Table IV. Same as Table III except for charmless
$B_{u,d}\to VP$ decays.}
\begin{center}
\begin{tabular}{l l c c c c c c c c l}
 &  & \multicolumn{4}{c}{$\nc(LL)=2$}
 &   \multicolumn{4}{c}{$\nc(LL)=\nc(LR)$}  \\ \cline{3-6} \cline{7-10}
\raisebox{2.0ex}[0cm][0cm]{Decay} &
\raisebox{2.0ex}[0cm][0cm]{Class} & 2 & 3 & 6 & $\infty$ & 2 & 3 &
6 & $\infty$  & \raisebox{2.0ex}[0cm][0cm]{Expt.}\\ \hline
\parbox[c]{2.8cm}{$\ov B^0_d \to \rho^-  \pi^+ $ \\
\\$\ov B^0_d \to \rho^+  \pi^- $\\} & \parbox{1cm}{I \\
\\${\rm I}$\\} &
\parbox[c]{1cm}{\centering 18.5\\ 26.5 \\ 14.8\\ 8.54 }&
\parbox[c]{1cm}{\centering 18.5 \\26.5\\14.6\\8.34 }
 & \parbox[c]{1cm}{\centering 18.5 \\ 26.5\\14.8\\8.43}
& \parbox[c]{1cm}{\centering 18.5 \\ 26.5\\14.9\\8.51} &
\parbox[c]{1cm}{\centering 18.5 \\ 26.5\\14.8\\8.54} &
\parbox[c]{1cm}{\centering 21.0 \\ 30.0\\16.3\\9.32} &
\parbox[c]{1cm}{\centering 23.6 \\ 33.8\\18.3\\10.4} &
\parbox[c]{1cm}{\centering 26.4 \\ 37.7\\20.3\\11.6} & $
\Bigg\}27.6^{+8.4}_{-7.4}\pm4.2$ \vspace{0.8mm}
\\
$\ov B^0_d \to\rho^+K^{-} $&      I,IV
&2.27&2.59&2.96&3.36&2.27&2.45&2.64&2.85&$ <25$\\
&&1.31&1.49&1.70&1.93&1.31&1.41&1.52&1.64& \\
 $\ov B^0_d \to \rho^0 \pi^0$&       II
&0.73&0.72&0.71&0.70&0.73&0.01&0.54&2.33&$<5.1$\\
&&0.80&0.79&0.78&0.77&0.80&0.02&0.50&2.22& \\
$\ov B^0_d \to \omega \pi^0$ &      II
&0.16&0.07&0.02&0.003&0.16&0.07&0.02&0.01&$<5.8$\\
&&0.19&0.10&0.05&0.04&0.19&0.09&0.07&0.14& \\
$ \ov B^0_d \to\omega \eta $& II
&0.32&0.35&0.39&0.44&0.32&0.004&0.28&1.14&$<12$\\
&&0.30&0.32&0.35&0.41&0.30&0.02&0.29&1.13& \\
$ \ov B^0_d \to\omega \eta'$& II
&0.30&0.28&0.25&0.23&0.30&0.04&0.15&0.66&$<60$\\
&&0.23&0.24&0.23&0.23&0.23&0.01&0.15&0.66& \\
$\ov B^0_d\to\rho^0\eta $&      II
&0.002&0.002&0.003&0.003&0.002&0.002&0.003&0.006&$<10$ \\
&&0.05&0.05&0.05&0.05&0.05&0.01&0.02&0.10& \\
 $ \ov B^0_d\to\rho^0 \eta'$&      II,VI
&0.04&0.01&0.01&0.05&0.04&0.05&0.06&0.08&$<23$ \\
&&0.02&0.02&0.04&0.08&0.02&0.02&0.05&0.12& \\
$ B^- \to \rho^-\pi^0$   &  III
&13.2&13.3&13.3&13.3&13.2&12.0&10.9&9.77&$<77$ \\
&&17.1&17.1&17.2&17.2&17.1&16.8&16.5&16.2& \\
$ B^- \to \rho^0\pi^- $  &  III
&13.1&13.2&13.3&13.4&13.1&9.98&7.29&5.04&$10.4^{+3.3}_{-3.4}\pm2.1$\\
&&9.64&9.69&9.76&9.82&9.64&6.14&3.46&1.61& \\
$ B^- \to \omega\pi^- $  &  III
&9.56&10.1&10.8&11.5&9.56&7.80&6.31&5.09&
$11.3^{+3.3}_{-2.9}\pm1.5$ \\
&&6.25&6.71&7.27&7.93&6.25&4.20&2.64&1.57& \\
$ B^- \to \rho^-\eta $  &   III
&9.46&9.70&10.0&10.3&9.46&7.95&6.62&5.41&$<15$ \\
&&11.9&11.8&12.1&12.3&11.9&10.8&10.0&9.25& \\
$ B^- \to \rho^-\eta'$  &   III
&5.49&4.88&4.45&4.20&5.49&4.27&3.21&2.31&$<47$ \\
&&6.97&6.43&6.00&5.67&6.97&6.16&5.40&4.69& \\
$ B^- \to\rho^0 K^-$&     III,VI
&0.86&0.90&0.97&1.05&0.86&0.63&0.46&0.30&$ <22$\\
&&0.57&0.56&0.57&0.59&0.57&0.34&0.18&0.07& \\
$\ov B^0_d \to K^{*-} \pi^+$ &       IV
&5.63&5.63&5.63&5.63&5.63&6.27&6.95&7.66&$22^{+8+4}_{-6-5}$\\
&&8.13&8.13&8.13&8.13&8.13&9.06&10.3&11.1& \\
$ B^- \to\ov K^{*0}\pi^- $&  IV
&3.41&3.41&3.41&3.41&3.41&4.34&5.38&6.54&$<27$\\
&&4.93&4.93&4.93&4.93&4.93&6.27&7.78&9.45& \\
$ B^-\to  K^{*0} K^-$&     IV
&0.38&0.39&0.39&0.39&0.38&0.49&0.61&0.74&$<12$\\
&&0.43&0.44&0.44&0.44&0.43&0.56&0.69&0.84& \\
$ B^- \to K^{*-}K^0$&         IV
&0.26&0.30&0.35&0.41&0.26&0.23&0.20&0.18&$-$\\
&&0.12&0.14&0.17&0.19&0.12&0.11&0.09&0.08& \\
\end{tabular}
\end{center}
\end{table}

\begin{table}[ht]
{\footnotesize Table IV. (Continued)}
\begin{center}
\begin{tabular}{l l c c c c c c c c l}
 &  & \multicolumn{4}{c}{$\nc(LL)=2$}
 &   \multicolumn{4}{c}{$\nc(LL)=\nc(LR)$}  \\ \cline{3-6} \cline{7-10}
\raisebox{2.0ex}[0cm][0cm]{Decay} &
\raisebox{2.0ex}[0cm][0cm]{Class} & 2 & 3 & 6 & $\infty$ & 2 & 3 &
6 & $\infty$  & \raisebox{2.0ex}[0cm][0cm]{Expt.}\\ \hline $ \ov
B^0_d \to \phi\pi^0$  &   V
&0.02&0.001&0.01&0.03&0.02&0.002&0.05&0.17&$<5.4$\\
&&0.03&0.002&0.01&0.04&0.03&0.003&0.08&0.25& \\
$ \ov B^0_d\to\phi \eta $ &   V
&0.01&0.001&0.003&0.02&0.01&0.001&0.03&0.09&$<9$\\
&&0.02&0.001&0.004&0.02&0.02&0.002&0.04&0.14& \\
$ \ov B^0_d\to\phi \eta'$  &  V
&0.006&0.0005&0.002&0.01&0.006&0.001&0.02&0.06&$<31$\\
&&0.01&0.001&0.002&0.01&0.01&0.001&0.03&0.09& \\
$ B^-\to\phi\pi^- $  &   V
&0.04&0.003&0.01&0.06&0.04&0.005&0.11&0.37&$<4.0$\\
&&0.06&0.005&0.01&0.09&0.06&0.01&0.17&0.54& \\
$\ov B^0_d\to K^{*0}\ov K^0$&  VI
&0.36&0.36&0.36&0.36&0.36&0.46&0.57&0.69&$-$\\
&&0.41&0.41&0.41&0.41&0.41&0.52&0.64&0.78& \\
$\ov B^0_d\to \ov K^{*0}  K^0$ &  VI
&0.24&0.28&0.33&0.38&0.24&0.21&0.19&0.16& $-$\\
&&0.11&0.13&0.15&0.18&0.11&0.10&0.09&0.08& \\
$\ov B^0_d\to\ov K^{*0}\pi^0$&     VI
&0.67&0.67&0.68&0.69&0.67&0.93&1.28&1.74&$<4.2$\\
&&1.43&1.45&1.45&1.46&1.43&1.93&2.52&3.21& \\
$\ov B^0_d\to\rho^0\ov K^0$&      VI
&2.92&3.32&3.68&4.07&2.92&2.74&2.58&2.50&$ <27$\\
&&2.23&2.50&2.75&3.01&2.23&2.10&2.01&2.00& \\
$\ov B^0_d \to\omega\ov K^0$&    VI
&0.33&0.16&1.55&4.55&0.33&0.71&5.01&13.3&$<21 $\\
&&0.83&0.04&0.89&3.45&0.83&0.34&4.37&13.0& \\
$\ov B^0_d \to \ov K^{*0} \eta$&  VI
&7.28&8,84&10.5&12.4&7.28&6.90&6.53&6.20&
$13.8^{+5.5}_{-4.6}\pm 1.6$\\
&&5.49&6.42&7.40&8.45&5.49&5.64&5.78&5.94& \\
$\ov B^0_d\to\ov K^{*0}\eta'$&    VI
&4.18&1.91&0.52&0.41&4.18&4.98&5.41&5.88&$<24$\\
&&1.10&0.43&0.19&0.56&1.10&1.28&1.30&1.33& \\
$\ov B^0_d \to\phi\ov K^0 $& VI
&11.3&7.15&3.97&1.74&11.3&5.87&2.22&0.32&$<28$\\
&&13.0&8.23&4.57&2.00&13.0&6.76&2.56&0.37& \\
$ B^- \to K^{*-}\pi^0$  &        VI
&4.46&4.45&4.43&4.42&4.46&4.68&4.92&5.17&$<800$\\
&&5.50&5.50&5.48&5.47&5.50&5.90&6.32&6.76& \\
$ B^- \to\rho^- \ov K^0$  &     VI
&2.53&2.99&3.49&4.02&2.53&2.27&2.02&1.78&  $<140$\\
&&1.46&1.72&2.01&2.32&1.46&1.31&1.16&1.03& \\
$ B^-\to \phi K^- $& VI
&11.6&7.71&4.28&1.87&11.6&6.33&2.39&0.35& $<5.9$\\
&&13.2&8.87&4.92&2.16&13.2&7.29&2.75&0.40& \\
$ B^- \to K^{*-}\eta$  &     VI
&9.60&11.4&13.3&15.4&9.60&8.51&7.50&6.57&
$26.4^{+9.6}_{-8.2}\pm 3.3$\\
&&7.93&9.01&10.2&11.4&7.93&7.45&6.99&6.55& \\
$ B^- \to K^{*-}\eta'$  &    VI,III
&4.93&2.16&0.76&0.71&4.93&5.61&6.36&7.18&$<35$\\
&&1.43&0.67&0.57&1.12&1.43&1.58&1.75&1.95& \\
$ B^-  \to \omega K^- $  &   VI,III
&1.50&0.78&1.79&4.53&1.50&1.04&5.30&14.3&$<8.0$\\
&&1.87&0.56&1.09&3.46&1.87&0.55&4.53&13.8&
\end{tabular}
\end{center}
\end{table}

\vskip 0.4cm
\begin{table}[ht]
{\footnotesize Table V. Same as Table III except for charmless
$B_{u,d}\to VV$ decays.}
\begin{center}
\begin{tabular}{l l c c c c c c c c l}
 &  & \multicolumn{4}{c}{$\nc(LL)=2$}
 &   \multicolumn{4}{c}{$\nc(LL)=\nc(LR)$}  \\ \cline{3-6} \cline{7-10}
\raisebox{2.0ex}[0cm][0cm]{Decay} &
\raisebox{2.0ex}[0cm][0cm]{Class} & 2 & 3 & 6 & $\infty$ & 2 & 3 &
6 & $\infty$  & \raisebox{2.0ex}[0cm][0cm]{Expt.}\\ \hline
 $\ov B^0_d \to \rho^-  \rho^+$   &I
  &32.0&32.0&32.0&32.0&32.0&36.1&40.5&45.1&  $ <2200$\\
 &&19.6&19.6&19.6&19.6&19.6&22.1&24.8&27.6&  \\
 $\ov B^0_d \to \rho^0 \rho^0$    &II
  &1.17&1.17&1.17&1.17&1.17&0.14&0.43&2.02&$<40$\\
 &&0.72&0.72&0.72&0.72&0.72&0.09&0.26&1.23& \\
 $ \ov B^0_d \to \omega\omega$    &II
  &0.71&0.58&0.54&0.60&0.71&0.20&0.53&1.73& $ <19$\\
 &&0.44&0.36&0.33&0.37&0.44&0.12&0.33&1.07& \\
 $B^-\to  \rho^- \rho^0$          &III
  &27.0&27.0&27.0&27.0&27.0&21.3&16.3&12.0& $<120$\\
 &&16.5&16.5&16.5&16.5&16.5&13.0&9.97&7.32& \\
 $ B^- \to \rho^-\omega$          &III
  &21.9&22.5&23.2&24.0&21.9&18.3&15.1&12.3& $<47$\\
 &&13.5&13.9&14.3&14.8&13.5&11.3&9.34&7.61& \\
 $\ov B^0_d \to K^{*-} \rho^+$    &IV
  &9.89&9.89&9.89&9.89&9.89&11.1&12.3&13.6&$ -$\\
 &&6.16&6.16&6.16&6.16&6.16&6.89&7.65&8.46& \\
 $\ov B^0_d \to\ov K^{*0}\rho^0$ &IV
  &1.00&0.99&0.98&0.96&1.00&1.35&1.89&2.62&$ <460 $ \\
 &&0.73&0.71&0.71&0.70&0.73&0.97&1.33&1.80& \\
 $\ov B^0_d \to\ov K^{*0}K^{*0}$ &IV
  &0.64&0.64&0.64&0.64&0.64&0.81&1.01&1.23& $-$\\
 &&0.32&0.32&0.32&0.32&0.32&0.41&0.51&0.62& \\
 $ B^- \to K^{*- }\rho^0$         &IV
  &9.98&10.0&10.1&10.1&9.98&10.4&10.9&11.5&$ <900$\\
 &&5.80&5.87&5.90&5.92&5.80&6.12&6.44&6.82& \\
 $ B^- \to \ov K^{*0} \rho^-$    &IV
  &6.42&6.42&6.42&6.42&6.42&8.20&10.2&12.4&$-$\\
 &&4.00&4.00&4.00&4.00&4.00&5.11&6.36&7.75& \\
 $ B^- \to K^{*- } K^{*0}$        &IV
  &0.68&0.68&0.68&0.68&0.68&0.87&1.08&1.31& $-$\\
 &&0.34&0.34&0.34&0.34&0.34&0.43&0.54&0.66& \\
 $ \ov B^0_d \to \rho^{0} \phi$   &V
  &0.03&0.002&0.01&0.06&0.03&0.01&0.10&0.33&  $ <13$\\
 &&0.02&0.001&0.006&0.04&0.02&0.003&0.07&0.21& \\
 $ \ov B^0_d \to\omega \phi$     &V
  &0.03&0.002&0.01&0.06&0.03&0.01&0.10&0.33& $ <21$\\
 &&0.02&0.001&0.006&0.04&0.02&0.003&0.07&0.21& \\
 $ B^- \to\rho^{-} \phi$          &V
  &0.07&0.004&0.02&0.12&0.07&0.01&0.22&0.70&  $ <16$\\
 &&0.05&0.003&0.01&0.08&0.05&0.01&0.14&0.44& \\
 $ \ov B^0_d \to \rho^0 \omega$   & VI
  &0.65&0.42&0.24&0.11&0.65&0.32&0.11&0.02& $ <11$\\
 &&0.40&0.26&0.15&0.07&0.40&0.20&0.07&0.01& \\
$\ov B^0_d \to \ov K^{*0}\omega$  &VI
  &15.0&7.80&2.91&0.43&15.0&4.54&0.19&2.01&$  <19$\\
 &&8.17&4.48&1.84&0.37&8.17&2.81&0.25&0.62& \\
 $ \ov B^0_d \to\ov K^{*0} \phi$ &VI
  &21.3&13.3&7.28&3.08&21.3&11.0&4.05&0.53&$<21$\\
 &&11.1&6.79&3.70&1.56&11.1&5.58&2.06&0.27& \\
 $ B^- \to K^{*-} \omega $        &VI
  &19.7&11.4&5.58&2.31&19.7&6.70&1.04&2.79&$ <52$\\
 &&10.9&6.64&3.48&1.58&10.9&4.15&0.83&1.09& \\
 $ B^- \to K^{*-} \phi$           &VI
  &22.7&14.2&7.74&3.27&22.7&11.7&4.31&0.57&$<41$\\
 &&11.8&7.22&3.94&1.66&11.8&5.93&2.19&0.29&
\end{tabular}
\end{center}
\end{table}


\newpage
\begin{thebibliography}{99}
\newcommand{\bi}{\bibitem}

\bi{Gao} CLEO Collaboration, Y.S. Gao and F. W\"urthwein,
hep-ex/9904008.

\bi{Bishai} CLEO Collaboration, M. Bishai {\it et al.,} CLEO CONF
99-13, hep-ex/9908018.

\bi{Smith} CLEO Collaboration, J.G. Smith, invited talk presented
at the Third International Conference on B Physics and CP
Violation, Taipei, Taiwan, Dec. 3-7, 1999.

\bi{Richichi} CLEO Collaboration, S.J. Richichi {\it et al.,}
hep-ex/9912059.

\bi{Kwon} CLEO Collaboration, D. Cronin-Hennessy {\it et al.,}
hep-ex/0001010.

\bi{Li} T.W. Yeh and H-n. Li, \pr {\bf D56}, 1615 (1997); H-n. Li
and B. Tseng, {\sl ibid.} {\bf D57}, 443 (1998).

\bi{Beneke} M. Beneke, G. Buchalla, M. Neubert, and C.T.
Sachrajda, \prl {\bf 83}, 1914 (1999).

\bi{CCTY} Y.H. Chen, H.Y. Cheng, B. Tseng, and K.C. Yang, \pr {\bf
D60}, 094014 (1999).

\bi{Buras98} A.J. Buras and L. Silvestrini, \np {\bf B548}, 293
(1999).

\bi{CLY} H.Y. Cheng, H-n. Li, and K.C. Yang, \pr  {\bf D60},
094005 (1999).

\bi{Ali} A. Ali and C. Greub, \pr {\bf D57}, 2996 (1998); A. Ali,
G. Kramer, and C.D. L\"u, \pr {\bf D58}, 094009 (1998).

\bi{CT98} H.Y. Cheng and B. Tseng, \pr {\bf D58}, 094005 (1998).

\bi{Cheng94} H.Y. Cheng, \pl {\bf B395}, 345 (1994); in {\it
Particle Theory and Phenomenology,}  XVII International Karimierz
Meeting on Particle Physics, Iowa State University, May 1995,
edited by K.E. Lassila {\it et al}. (World Scientific, Singapore,
1996), p.122\,.

\bi{Kamal94} A.N. Kamal and A.B. Santra, Alberta Thy-31-94 (1994);
\zp {\bf C72}, 91 (1996).

\bibitem{Soares}
J.M. Soares, \pr {\bf D51}, 3518 (1995).

\bi{Agashe} K. Agashe and N.G. Deshpande, hep-ph/9909298.

\bi{LEP} D. Abbaneo {\it et al.,} LEPVUB-99-01, 1999.

\bi{Behrens2} CLEO Collaboration, B.H. Behrens {\it et al.,}
hep-ex/9905056.

\bi{BSW} M. Wirbel, B. Stech, and M. Bauer, {\sl ibid.} {\bf C29},
637 (1985); M. Bauer, B. Stech, and M. Wirbel, \zp {\bf C34}, 103
(1987).

\bi{gamma} X.G. He, W.S. Hou, and K.C. Yang, \prl {\bf 83}, 1100
(1999).

\bi{Gerard} J.-M. G\'erard, J. Pestieau, and J. Weyers, \pl {\bf
B436}, 363 (1998).

\bi{Martinelli} G. Martinelli, invited talk presented at Kaon '99,
Chicago, June 21-26, 1999.

\bi{CLEOpiK} CLEO Collaboration, R. Godang {\it et al.,} \prl {\bf
80}, 3456 (1998).

\bi{Deshpande} N.G. Deshpande, X.G. He, W.S. Hou, and S. Pakvasa,
\prl {\bf 82}, 2240 (1999).

\bi{Ball} P. Ball and V.M. Braun, \pr {\bf D58}, 094016 (1998); P.
Ball, {\sl J. High Energy Phys.} {\bf 9809}, 005 (1998)
[hep-ph/9802394].

\bi{Hou} W.S. Hou, J.G. Smith, and F. W\"urthwein, hep-ph/9910014.

\bi{CY} H.Y. Cheng and K.C. Yang, \pr {\bf D59}, 092004 (1999).

\bi{Behrens} CLEO Collaboration, B.H. Behrens {\it et al.,} \prl
{\bf 80}, 3710 (1998).

\bi{Halperin} I. Halperin and A. Zhitnitsky, \pr {\bf D56}, 7247
(1997); E.V. Shuryak and A. Zhitnitsky, \pr {\bf D57}, 2001
(1998).

\bi{Ahmady} M.R. Ahmady, E. Kou, and A. Sugamoto, \pr {\bf D58},
014015 (1998).

\bi{Du} D.S. Du, C.S. Kim, and Y.D. Yang, \pl {\bf B419}, 369
(1998); D.S. Du, Y.D. Yang, and G. Zhu, \pr {\bf D59}, 014007
(1999); {\sl ibid.} {\bf D60}, 054015 (1999).


\bi{CLEOomega} CLEO Collaboration, T. Bergfeld {\it et al.,} \prl
{\bf 81}, 272 (1998).

\bi{Gronau} M. Gronau and J.L. Rosner, hep-ph/9909478.

\end{thebibliography}
\end{document}